\documentstyle[preprint,prl,aps,epsfig]{revtex}
\tightenlines
\begin{document}

\draft

\title{
Spin susceptibility of neutron matter at zero temperature
}

\author{
S. Fantoni$^{1,2}$, A. Sarsa$^1$ and K. E. Schmidt$^3$
}

\address{
$^1$ International School for Advanced Studies, SISSA, 
I-{\rm 34014} Trieste, Italy\\
$^2$ International Centre for Theoretical Physics, ICTP, 
I-{\rm 34014} Trieste, Italy\\
$^3$ Department of Physics and Astronomy, Arizona State University,
Tempe, AZ {\rm 85287} U.S.A. \\}

\maketitle

\begin{abstract}

The Auxiliary Field Diffusion Monte Carlo method is applied to 
compute the spin susceptibility and the compressibility of neutron matter
at zero temperature.
Results are given for realistic interactions which include both
a two--body potential of the Argonne type  
and the Urbana IX three--body potential.
Simulations have been carried out for about 60 neutrons.
We find an overall reduction of the spin susceptibilty by
about a factor of 3 with respect to the Pauli susceptibility for
a wide range of densities. Results for the compressibility
of neutron matter are also presented and compared with other available
estimates obtained for semirealistic nucleon--nucleon interactions and
with more traditional many--body techniques, like Brueckner's or 
Correlated Basis Function theories. 

\end{abstract}

\pacs{PACS numbers: 26.60.+c, 26.50.+x, 05.10.Ln}

In this paper we show that the strong correlations induced by realistic
nucleon--nucleon interactions reduce, by up to about a factor of 3, the spin
susceptibility, $\chi$, of degenerate neutron matter. 
This reduction may have important implications for problems of 
astrophysical interest, like for
instance, neutrino scattering rates in dense matter
and, more generally, the study of supernovae and 
proto--neutron stars\cite{raffelt96,reddy99}.  

Starting from the pioneering work by Sawyer\cite{sawyer75},
several calculations of the neutrino mean free path in uniform nuclear matter
have been performed\cite{reddy99,iwamoto82,friman79,burrows99}
which show that the effects due to strong interactions are relevant, 
particularly in the spin-density channel which couples with the 
axial vector current. A sizable reduction of $\chi$, or, equivalently,
a large value of the $G_0$ Landau parameter leads to an appreciable 
suppression of the Gamow--Teller transitions. 

The neutrino momenta and the momentum transfers in many applications are
small compared with the neutron Fermi momentum, and both energy transfers
and the temperature are small compared with the neutron Fermi Energy. 
Therefore, the Landau parameters of neutron matter at zero temperature
are the main quantities to compute in order to evaluate the mean free path
of a neutrino in dense matter.

In this respect, {\sl ab initio} calculations
of the Landau parameters or related quantities, such as the compressibility
${\cal K}$, the effective mass $m^{\ast}$ or the spin susceptibility $\chi$, 
for degenerate neutron matter are extremely important. Such calculations
can now be performed 
because of recent advances in many--body methods, particularly those
based on quantum simulations, and because of 
the much improved knowledge of the nucleon--nucleon interaction. 

Previous evaluations of the Landau parameters $F_l$ and $G_l$ (with $l \leq 1$)
for neutron matter were based either on Skyrme--type potential 
models\cite{reddy99} or on microscopic calculations performed with semirealistic
bare interactions\cite{clark69,backmann73,jackson92}. 
The qualitative behaviour of the compressibility ratio is 
\begin{eqnarray}
\frac{\cal K}{{\cal K}_F} = \frac{1+\frac{1}{3} F_1}{1+F_0}\ , 
\label{comp_ratio}
\end{eqnarray}
where ${\cal K}_F$ the 
compressibility of the non interacting Fermi gas, is similar in most of 
the various calculations. On the contrary, the spin susceptibility ratio, 
which is approximately related to $G_0$ by 
\begin{eqnarray}
\frac{\chi}{\chi_F} \approx \frac{1+\frac{1}{3} F_1}{1+G_0}\ , 
\label{susc_ratio}
\end{eqnarray}
may differ up to a factor of 3 at equilibrium density of nuclear matter 
$\rho_0=0.16~ {\rm fm}^{-3}$, and even more at higher densities.

The Skyrme models generally predict much smaller 
values of $G_0$ with respect to
microscopic calculations, and become unstable to spin oscillations, driving
toward  ferromagnetic ground states\cite{reddy99}, 
since $1+G_0$ becomes negative for
densities in the range $(2-4)\rho_0$. At small momentum transfer, 
N--N correlations enhance
the spin--response approximately by a factor $(\chi/\chi_F)^2$. 
Therefore,  Skyrme models
predict neutrino mean free paths smaller than microscopic 
models.

On the other hand, the existing microscopic calculations have been
performed with old semirealistic interactions, like the Reid or the 
Bethe--Johnston potentials, and no three--body force. In addition, the
many--body methods used until now may be questionable 
for the convergence of the underlying perturbation theory as well as
for the treatment of spin--dependent correlations.

We report the results of quantum simulations of neutron matter
for the old Reid--$v_6$ potential \cite{reid} and modern 
realistic interactions, based upon the Argonne $v_{18}$ two--body
potential plus the Urbana three--body potential, UIX,
\cite{pieper98}, denoted hereafter as AU18. It is well known that such
interactions provide a realistic description of light nuclei and nuclear 
matter \cite{wiringa,apr98}. 

The quantum simulations have been carried out by using 
a new Diffusion Monte Carlo 
method, the so called AFDMC method\cite{schmidt99}. It introduces  
auxiliary field variables in the simulation, which makes it   
the only existing quantum Monte Carlo method that
can handle spin--dependent nuclear Hamiltonians and
a relatively large number of nucleons. 
In this approach the scalar parts of the Hamiltonian are propagated as in 
standard Diffusion Monte Carlo (DMC) \cite{gfmc84}.
Auxiliary fields are introduced
to replace the spin--isospin dependent interactions between pairs of particles
with interactions between particles and auxiliary fields. Integrating
over the auxiliary fields reproduces the original spin--isospin
dependent interaction. The method consists of a Monte Carlo sampling of the
auxiliary fields and then propagating the spin--isospin variables at the
sampled values of the auxiliary fields. This propagation results in a
rotation of each particle's spin--isospin spinor. 

The guiding function, $\Psi_T$, in our AFDMC calculation is a simple 
trial function
given by a Slater determinant of one--body space--spin orbitals multiplied
by a central Jastrow correlation. 
The orbitals are plane waves that
fit in the box times two component spinors, corresponding to neutron--up and
neutron--down states. 
The overlap of a walker with this wave function is the determinant of the
space--spin orbitals, evaluated at the walker position and spinor for 
each particle, and multiplied by the scalar Jastrow product.
Such an overlap is complex, 
so the usual fermion sign problem becomes a phase problem. We constrain the
path of the walkers to a region where the real part of the overlap with our
trial function is positive. For spin--independent potentials this reduces
to the fixed--node approximation. 

This method has already been applied to unpolarized neutron matter 
and neutron drops ($A=7,8$) with fairly realistic interactions
that include tensor, spin--orbit and three--body terms. The 
neutron matter calculations have 
been done with up to 66 neutrons in a periodic box with a low
variance ($<0.1 $ MeV per nucleon). The calculation
scales in particle number roughly like fermion Monte Carlo 
with central forces \cite{FSS00,schmidt01}.

We compute the spin susceptibility by applying a magnetic field to the 
system. Ignoring any orbital effects, the Hamiltonian is given by
\begin{equation}
H = H_0 - \sum_i \vec \sigma_i \cdot \vec b \ ,
\end{equation}
where $\vec{b}=\mu~\vec{B}$ and 
$\mu= 6.030774 \times 10^{-18} {\rm MeV/Gauss}$, and
the susceptibility is defined as
\begin{equation}
\label{susdef}
\chi =\left . - \rho \mu^2
\frac{\partial^2 E_0(b)}{\partial b^2} \right |_{b=0}\ ,
\end{equation}
where $\rho$ is the number density and $E_0(b)$ is the ground energy
in field $b$.

Let us use the Pauli expansion of the energy per particle as 
a function of the spin  polarization 
$p=-\partial E_0(b)/ \partial b|_{b=0}$

\begin{equation}
\label{epauli}
E(p) = E(0) - b ~ p + \frac{1}{2} p^2 E''(0) \ ,
\end{equation}

where the derivatives are with respect to the
polarization.
Minimizing $E(p)$ with respect to $p$ one gets the following result for the
spin susceptibility 

\begin{equation}
\label{suspauli}
\chi = \mu^2 \rho \frac{1}{E''(0)} \ .
\end{equation}

For a noninteracting Fermi gas the spin susceptibility is 
$\chi_F =  \mu^2 m k_f/(\hbar^2 \pi^2)$. 
AFDMC allows us to get the energy eigenvalue, $E_0(J_z,b)$, for 
the interacting system in a field $b$ for a state of a given spin asymmetry
$J_z=N_{\uparrow}-N_{\downarrow}$.

Assuming that the energy and polarization are known in terms of $J_z$, 
$E''(0)$ in Eq. (\ref{suspauli}) can be obtained as a 
straightforward application of the chain rule 

\begin{equation}
E''(0) =
\left[ \frac{\partial p}{\partial J_z} \right ]^{-2}
\left \{
\frac{\partial^2 E_0}{\partial J_z^2}
-
\frac{\partial E_0}{\partial J_z}
\left[ \frac{\partial p}{\partial J_z} \right ]^{-1}
\frac{\partial^2 p}{\partial J_z^2}
\right \} \ .
\end{equation}
Since we are calculating the lowest energy state, the derivative
of the energy with respect to $J_z$ vanishes. Therefore this result
reduces to
\begin{equation}
E''(0) =
\left[ \frac{\partial p}{\partial J_z} \right ]^{-2}
\frac{\partial^2 E_0}{\partial J_z^2} \ .
\label{esecond}
\end{equation}

\begin{figure}
\caption{
The energy of non interacting neutrons as a function of magnetic field
$b$ at $\rho = 2\rho_0$
for various finite-sized closed shell trial functions with spin up and
down values shown. Also plotted is the correct infinite system energy
and the parabolic Pauli value. 
}
\vspace{0.4cm}
\hspace{-1.2cm}
\label{f1}
\epsfxsize=36pc
\epsfbox{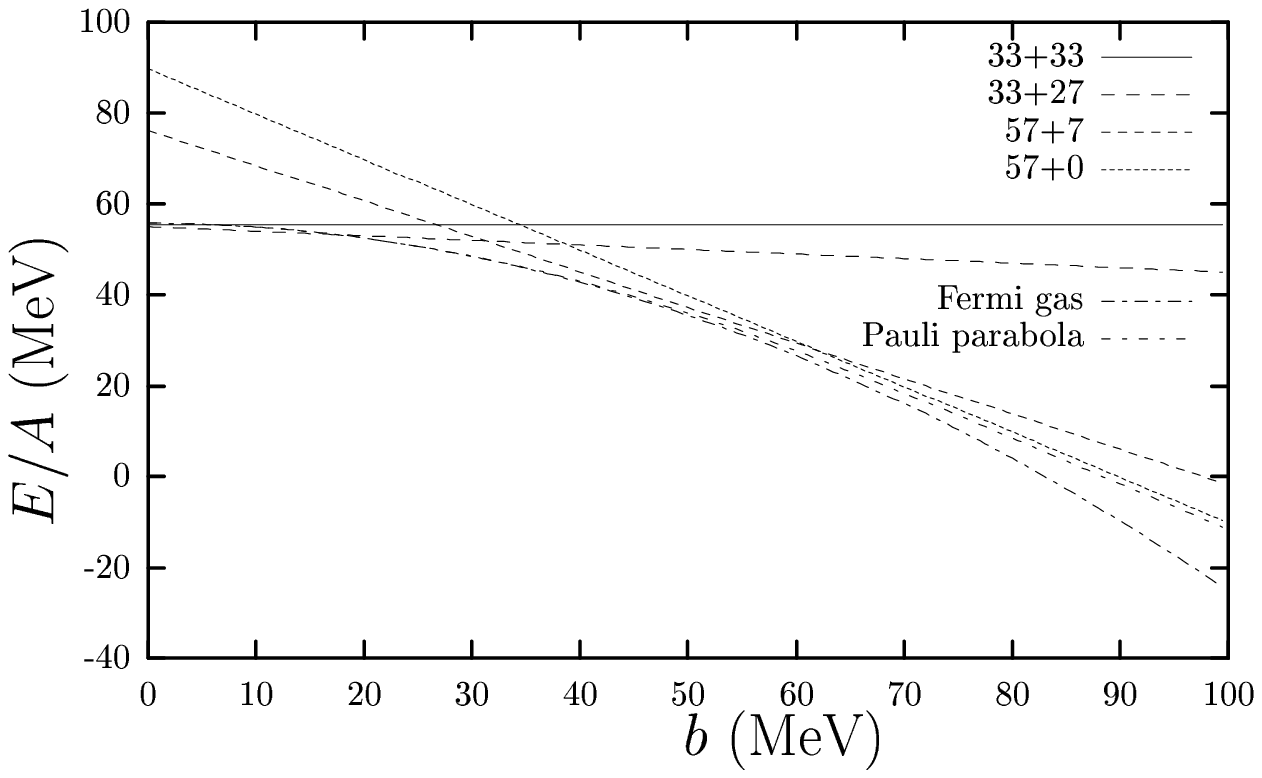}
\end{figure}

Let us consider the non interacting finite systems as a guide for 
the quantum simulations of the interacting ones. For such systems
the energy is not a quadratic function of
the external field $b$. In figure \ref{f1} we plot $E_0(J_z,b)$, as
a function of $b$, for four different 
systems with a finite number $N\sim 60$ of noninteracting neutrons in a 
periodic box at $\rho = 2\rho_0$. 
The cases, for which we have 
done simulations, are shown in the figure,
namely $(N_\uparrow,N_\downarrow)$=$(33,33)$, 
$(33,27)$, $(57,7)$ and $(57,0)$.  
One can see that the various $E(J_z,b)$ are linear in $b$ and each of them
is tangent to the Pauli parabola (which refers to infinite Fermi gas case) 
at some value $b_0$ of the field $b$.
We also compare in the figure the
Pauli parabola with the 
exact result for the Fermi gas at that density. They are very close up to
$b\sim 50$ MeV. Using Eq. (\ref{esecond}), 
one gets $\chi/\chi_F\sim 1$ for $J_z=50$ and
$J_z=57$.

In the interacting case, 
the derivatives in Eq. (\ref{esecond}) can 
be easily estimated by computing $E_0(J_z,b)$ with AFDMC,  and 
using the following equations

\begin{eqnarray}
\frac{\partial p}{\partial J_z} &\approx& 
\frac{E_0(J_z=J_{z0},b=0)-E_0(J_z=J_{z0},b=b_0)}
 {J_{z0}~b_0} \ , \\
\frac{\partial^2 E_0}{\partial J_z^2} &\approx& 2~
\frac{E_0(J_z=J_{z0},b=0)-E_0(J_z=0,b=0)}
{J_{z0}^2}\ , 
\end{eqnarray}
whose validity relies on
the following reasonable assumptions:
(i) for $b=0$, $E_0(J_z,b)$ is quadratic in $J_z$;
(ii) for a fixed $J_z$, $E_0(J_z,b)$ is linear in $b$;
(iii) the polarization is linear in $J_z$. These assumptions become exact
in the limit of an infinite system with $J_z$ and $b$ small.

The non--interacting case indicates the use of $J_{z0}=50$ 
and the value for $b_0$ 
at which $E(J_{z0}=50,b)$ is tangent to the 
Pauli parabola (for the 
density $\rho=2\rho_0$ of Fig. \ref{f1}, $b_0=53$ MeV).
Most of our calculations have been carried out with $J_{z0}=50$. 
At $\rho=1.25 \rho_0$ we have verified the linearity of
$E(J_{z0}=50,b)$ on $b$ beyond $b_0(1.25 \rho_0)=39 $MeV 
finding that the result
for $\chi/\chi_F$ is largely independent on the value of $b_0$.
We have also verified the dependence of  
$\chi/\chi_F$ on $J_{z0}$, 
performing simulations with $J_{z0}=6$ and
found that it is very weak. 

A time step $\Delta\tau=5 \times 10^{-5} {\rm MeV}^{-1}$ 
was sufficient in most of the quantum simulations to
obtain agreement between the mixed and the growth energies \cite{gfmc84}
within the statistical accuracy.

We have made simulations with the Reid $v_6$ interaction (Reid6), the same used
in the Correlated Basis Function calculation of Ref.\cite{jackson92}. Our
result for $\chi/\chi_F$, at $\rho=1.25\rho_0$, is about $25\%$ smaller, which
indicates that the CBF perturbative calculations of 
Ref.\cite{jackson92}, based on Jastrow--correlated basis functions,  
either have not reached a satisfactory convergence for the Landau
parameter $G_0$ or Eq (\ref{susc_ratio}) may not be sufficiently adequate
for the Reid6 interaction.

We have also considered realistic interactions, characterized by the so called
$v_6^{\prime}$ or $v_8^{\prime}$ two--body potentials plus UIX
three--body potential, and hereafter denoted as AU6' and AU8' respectively. 
The $v_8^{\prime}$ 
is a two--body potential of the Argonne type which includes the four central 
spin--isospin components, plus the four tensor 
and spin--orbit ones. It fits the nucleon--nucleon experimental data and
embodies the main features of the  Argonne $v_{18}$\cite{smerzi97}. 
The $v_6^{\prime}$
potential is the $v_8^{\prime}$ with its spin--orbit components removed.

Our results for $\chi/\chi_F$, obtained with the Reid6, 
AU6' and AU8' interactions 
are given in Table \ref{t1} and compared with previous
microscopic calcuations of the same quantity.
The Brueckner theory calculations of Ref.\cite{backmann73} 
are performed with the 
full Reid potential, and, therefore are not directly comparable with our
Reid6 calculations. However, we do not find a sizable contribution to
$\chi/\chi_F$ coming from the spin--orbit component of the two--body
potential. Our results for AU6' and AU8' coincide within the
statistical error. To address the problem of the influence of the three-body
force on the results 
we have calculated the spin susceptibility at $\rho=2\rho_0$
using the Argonne $v_6^{\prime}$ potential and no three body interactions. 
We have found that while 
the energy is reduced by roughly 25 MeV per particle, the spin
susceptibility ratio is practically unchanged (0.31(1)).

We have also calculated the compressibility ${\cal K}$, given by
\begin{equation}
\frac{1}{\cal K}=\rho^3~ \frac{\partial^2 E_0(\rho)}{\partial \rho^2}+
2\rho^2 \frac{\partial E_0(\rho)}{\partial \rho} \ ,
\end{equation}
where $E_0(\rho)$ is a polynomial fit to the AFDMC energies $E(J_z=0,b=0)$.
For a Fermi gas the compressibility is 
${\cal K}_F = 9 \pi^2 m /(k_f^5 \hbar^2)$. 
The AFDMC results for ${\cal K}/{\cal K}_F$, obtained with
the AU6' interaction are shown in 
Table \ref{t2}, where they are also compared with the corresponding
CBF estimates (AU6'-CBF) and other existing microscopic
calculations\cite{backmann73,jackson92,apr98}. 

\begin{figure}
\caption {
AFDMC equation of state of the AU6' model of neutron matter 
(dots); 
CBF theory [19] results for the same interaction model
are in the shaded area where the highest values correspond to the variational
estimate. The equation of state obtained
in Ref.[13] for the 
AU18 interaction by using FHNC/SOC theory is given by dashes. The errors
are smaller than the symbols.
}
\vspace{0.4cm}
\hspace{-1.2cm}
\epsfxsize=36pc
\epsfbox{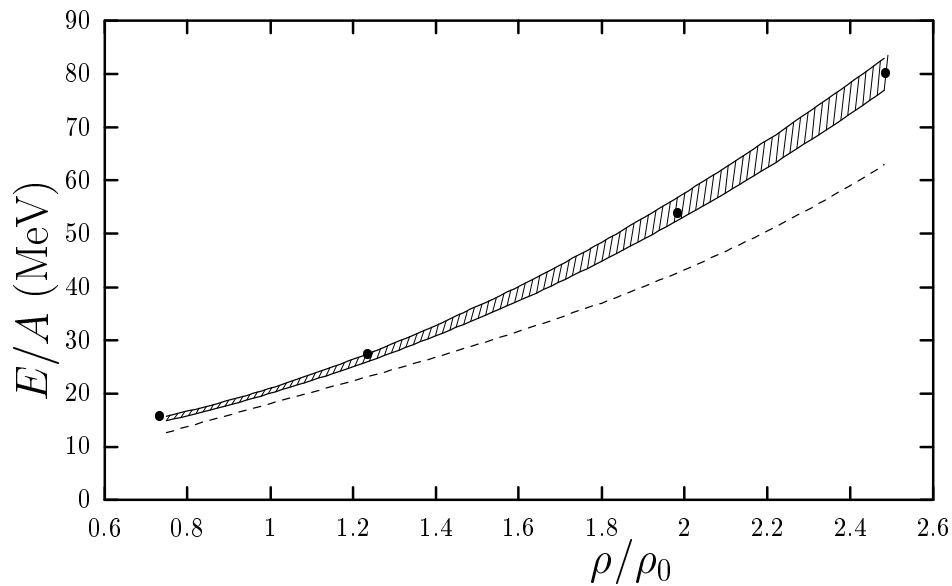}
\label{f2}
\end{figure}

For the sake of completeness we show in Fig. \ref{f2} the equations of state
of neutron matter, which have been used to compute the compressibility 
ratio of Table \ref{t2}.

The AU18 results are taken from 
Ref.\cite{apr98}, and have been obtained with the full AU18 model interaction,
by using variational FHNC/SOC methods.
The AU6'-CBF\cite{fabrocini01} results have been obtained by using 
the AU6' interaction, 
as in the AFDMC simulations, and essentially the same many--body technique 
as in Ref.\cite{apr98}. They
also include the corrections coming from the
lowest order elementary diagram, as discussed in 
Refs. \cite{manchester,schmidt01}. 

One can see that the AFDMC results for both the equation of state and the
compressibility of the AU6' model of neutron matter are in reasonably
good agreement with those obtained by using CBF theory.

We have estimated the finite size effects in the AFDMC simulations 
by performing variational calculations with the Periodic Box FHNC method of 
Ref.\cite{fs-pbox}. They indicate a correction which is at most $10\%$ of the
mixed energy per particle. An error of the same size is expected for the
compressibilty ratio and of an order of magnitude smaller for the
spin susceptibility.
A more detailed discussion of the AFDMC calculation and the 
equation of state of neutron matter will be given elsewhere\cite{schmidt01}. 

In conclusion, we have presented new results on the spin susceptibility
and the compressibility of neutron matter at zero temperature, which show
a strong reduction of these quantities with respect to their Fermi gas
values. The calculations have been performed for realistic model 
interactions, which include tensor, spin--orbit and
three--body terms, and by using the AFDMC method, a newly developed
quantum simulation technique, capable of dealing with strongly
spin--dependent interactions. 
Of particular relevance is the quenching by about a factor of 3  
of the spin susceptibility with respect
to the commonly used Pauli value, found at all
the various densities considered.  
Such a reduction has strong effects on the 
mean free path of a neutrino in dense matter and should be seriously taken into
account in the studies of supernovae and proto--neutron stars.

\section*{Acknowledgements}

We are indebted to A. Fabrocini for providing us the results of the 
second order perturbative corrections to the energy per particle of
neutron matter for the AU6' interaction and to C.J. Pethick for
insightful discussions.
Portions of this work were supported by MURST-National Research Projects
and CINECA computing center.

\begin{table}
\caption{ 
Spin susceptibility ratio $\chi/\chi_F$
of neutron matter. Our AFDMC results for the interactions AU6', AU8' and
Reid6 are compared with those obtained from 
Refs. [8,9] by using Eq(\ref{susc_ratio}).
The statistical error is given in  parentheses.
} 
\label{t1}
\begin{tabular}{c|cc|ccc}
 $\rho/\rho_0$ & Reid\cite{backmann73} & Reid6\cite{jackson92}  
 & AU6' & AU8'  & Reid6\\
\tableline
   $0.75$      & 0.45  & 0.53   & 0.40(1) & &  \\     
   $1.25$      & 0.42  & 0.50   & 0.37(1) & 0.39(1) & 0.36(1)  \\ 
   $2.0$       & 0.39  & 0.47   & 0.33(1) & 0.35(1) & \\        
   $2.5$       & 0.38  & 0.44   & 0.30(1) & &  \\
\end{tabular}
\end{table}

\begin{table}
\caption {
Compressibility ratio ${\cal K}/{\cal K}_F$ of neutron matter.
Our AFDMC results for the AU6' interaction are compared with those obtained
with CBF theory [19] (AU6'-CBF) and those of Refs. [8,9,13]
The statistical error is given in parentheses.
} 
\label{t2}
\begin{tabular}{c|ccc|cc}
 $\rho/\rho_0$ & Reid\cite{backmann73}   &  Reid6\cite{jackson92}   
 & AU18\cite{apr98} &  AU6'-CBF & AU6'  \\
\hline 
   $0.75$ & 0.91   & 2.06   & 1.10  & 0.85  & 0.89(3)  \\     
   $1.25$ & 0.70   & 1.35   & 0.71  & 0.45  & 0.47(3)  \\   
   $2.0$  & 0.49   & 0.77   & 0.26  & 0.23  & 0.21(3)  \\        
   $2.5$  & 0.42   & 0.60   & 0.15  & 0.17  & 0.14(3)  \\ 
\end{tabular}
\end{table}

\end{document}